\documentclass[12pt]{iopart}
\usepackage[utf8]{inputenc}
\usepackage[english]{babel}

\usepackage{soul}
\usepackage{amsmath}
\usepackage{graphicx}
\usepackage[square,sort,comma,numbers]{natbib}
\usepackage[arrowdel]{physics}
\usepackage{quantikz}
\usepackage{tikz}
\usetikzlibrary{quantikz}
\usepackage{url}

\begin{document}

\title[Revisiting semiconductor bulk Hamiltonians using quantum computers]{Revisiting semiconductor bulk hamiltonians using quantum computers}

\author{Raphael César de Souza Pimenta$^1$ \& Anibal Thiago Bezerra$^2$}

\address{$^1$ Departamento de Física, Universidade Federal de Santa Catarina,  Roberto Sampaio Gonzaga s/n, Florianópolis-SC, Brasil}

\address{$^2$ Departamento de Física, Universidade Federal de Alfenas, Jovino Fernandes Sales 2600, Alfenas-MG, Brazil}

\ead{anibal.bezerra@unifal-mg.edu.br}
\vspace{10pt}
\begin{indented}
\item[]August 2022
\end{indented}

\begin{abstract}
With the advent of near-term quantum computers, the simulation of properties of solids using quantum algorithms becomes possible. By an adequate description of the system's Hamiltonian, variational methods enable to fetch the band structure and other fundamental properties as transition probabilities. Here, we use k$\cdot$p Hamiltonians to describe semiconductor structures of the III-V family and obtain their band structures using a state vector solver, a probabilistic simulator, and a real noisy-device simulator. The resulting band structures are in good agreement with the ones obtained by direct diagonalization of the Hamiltonian. Simulation times depend on the optimizer, circuit depth, and simulator used. Finally, with the optimized eigenstates, we convey the inter-band absorption probability, demonstrating the possibility of analyzing the fundamental properties of crystalline systems using quantum computers.  

\end{abstract}

\vspace{2pc}
\noindent{\it Keywords}: {Quantum Computing, Semiconductor Band Structure, k$\cdot$p method}

\section{INTRODUCTION}





Materials science has been developed over the last decades with the great help of digital simulations on supercomputers. Such development enabled for several new applications by manipulating the structures, compositions, and atomic arrangement, for a broad range of materials~\cite{kreider2022thin}. We observe a change of paradigm when the scientific community starts primarily simulating the material's response, predicting the desired behaviors, and subsequently moving to the lab to concretely produce new devices with outstanding properties~\cite{kreider2022thin}. The screening of materials' and structures' properties is highly bound to the balance between the systems we can simulate and the computational time invested in the modeling~\cite{rahm2020library}. However, there are known limits to how far digital simulations can reach. The complexity of the simulation increases exponentially with the number of constituents of the system, and with the quantum correlations between electrons and nuclei~\cite{daley2022practical}. Hence, approximations should be employed in the models to ease the complexity. The density functional theory (DFT) is an example of success in the materials science area still at the cost of time-consuming simulations~\cite{kreider2022thin}. The rhythm to which computational power is increased seems not enough to follow the necessities of simulating new systems with desired properties~\cite{rahm2020library}. Therefore, the question is if there are viable alternatives to burden the barriers imposed by the limited computational power we have access to perform the simulations. 

Quantum computers appeared as a powerful tool to tackle complicated computational problems~\cite{daley2022practical,ippoliti2021many}. By performing computation using quantum states and their superposition, quantum computers are being employed to scale down the complexity of non-polynomial-hard problems, such as the decomposition of big numbers~\cite{shor1999polynomial}, obtaining optimized routes~\cite{srinivasan2018efficient}, and coloring graphs~\cite{verteletskyi2020measurement}. Quantum computers available today are the so-called Noise Intermediate Scale Quantum computers (NISQ)~\cite{preskill2018}, where the development of algorithms generally involves programming at the gate level, with gates subject to depolarization~\cite{bassman2022arqtic}. Although a plethora of algorithms have been proposed to scale down the complexity of computational problems using NISQ devices, for those problems related to solid state physics, the quantum advantage of quantum computers is still a promise~\cite{vorwerk2022quantum}, with a few exceptions, such as for two-dimensional out-of-equilibrium dynamics~\cite{daley2022practical}. To promote the quantum advantage, fault-tolerant gate-based quantum computers are required, emphasizing the importance of understanding and using NISQ devices.  



Since quantum computer states could mimic the physical behavior of microscopic systems, their use to obtain the electronic properties of solids seems to be the most natural application of quantum computing~\cite{daley2022practical}. The major difficulty is determining how to map the system on a quantum device. Some attempts to a fully \textit{ab-initio} simulation have been proposed by employing wave-function-based Hamiltonian descriptions, such as coupled clusters with single and double excitations (CCSD) theories. Some advance has been shown with periodic hydrogen chains~\cite{yoshioka2022,manrique2020} using variational algorithms and unitary coupled clusters (UCSSD) to evaluate both ground states~\cite{liu2020} and excited states~\cite{yoshioka2022}. Symmetries have been shown to reduce the number of terms of the Hamiltonian, allowing the use of NISQ computers~\cite{Kivlichan2018}. Still, for hydrogenoid chains, Adaptive Derivative-Assembled Pseudo-Trotter (ADAPT)~\cite{liu2020}, and other bases as Slater-type orbitals (STO-3G)~\cite{mizuta2021} ansatzes have been shown as alternatives to UCSSD description, yet adequate to be used for systems with periodic boundary conditions. 


Tight-binding models also play a crucial role in the endeavor of simulating the electronic properties of crystalline solids using NISQ computers. Sherbert and co-workers~\cite{sherbert2021systematic} propose an algorithm to evaluate the band structure of a simple cubic lattice structure by implementing the variational quantum eigensolver (VQE), the variational quantum deflation (VQD), and quantum phase estimation (QPE), obtaining both the ground and excited states sampled across a high-symmetry path in the Brillouin zone. The authors demonstrate results from \textit{s} and \textit{p} orbitals contribution, using both quantum simulators and real quantum hardware, applying error mitigation. For the quantum device simulator, error mitigation seems enough to help reproduce the expected band structure. However, for the real hardware, the mitigation routine returned worse results. Likewise, Cerasoli and co-workers~\cite{Cerasoli2020} evaluate the silicon band structure using a tight-binding Hamiltonian with VQE and VQD. The authors elucidate the influence of the classical optimization method and emphasize the importance of error analysis and mitigation when dealing with noisy devices. For a more general picture, Bassman and co-workers ~\cite{bassman2021} reviewed the use of quantum computers to evaluate the electronic properties of crystalline solids.

In the present work, we follow a different route from purely \textit{ab-initio} and tight-binding Hamiltonians. We employ a k$\cdot$p Hamiltonian to represent the crystalline structure of bulk semiconductors from the III-V family. The Hamiltonian is decomposed in terms of Pauli strings, whose expectation values can be directly evaluated using a quantum computer. We then bring a benchmark on the efficiency of the SSVQE algorithm to determine the band structure for a class of crystalline structures using quantum devices. We use the PeenyLane library to implement the hybrid algorithm using a state vector solver, a probabilistic simulator, and a quantum machine simulator. The resulting band structures are compared with the outcome of numerical diagonalizations. Additionally, we investigate the influence of different configurations in the simulation, such as the number of layers of the ansatz, and the optimizers for the classical optimization routine. The results are evaluated by comparing the time and cycles needed for the algorithm to converge. Finally, we use the optimized ansatz to determine the interband transition rate of several semiconductor compounds.

\section{METHODS}

\subsection{The $k\cdot p$ method}

Since its development, the $k\cdot p$ method has been widely used to analyze the electronic structure of semiconductors in the vicinity of high-symmetry points of the Brillouin zone~\cite{ben2003band,gawarecki2022invariant,neffati2012full}. The method is grounded on the influence of the crystalline periodic potential profile over the electron in the structure, taking into account the effective mass approximation~\cite{schubert2015physical,bastard1990wave}. For diamond-like structures, the electronic dispersion around the high-symmetry $\Gamma$-point ($\vec k = 2\pi(0,0,0)/a$) conveys the relevant knowledge pertaining to the optical and transport properties. Using the translational symmetry of the periodic lattice, together with the Block theorem, one can expand the Hamiltonian of the structure in series keeping only relevant contributions. The Schröedinger equation on the basis of the Bloch states $\ket{q\vec k}$, for wave vectors in the vicinity of the center of the Brillouin zone, reads~\cite{bastard1990wave}

\begin{equation}\label{eq:KP_Hamiltonian}
    \sum_l \left[\left(\varepsilon_{q\Gamma}-\varepsilon_{q\vec k}+\frac{\hbar^2 k^2}{2m_0}\delta_{ql}+ \frac{\hbar \vec k}{2m_0}\cdot \mel{q\Gamma}{\vec p + \frac{\hbar}{4m_0 c^2}(\vec \sigma \times \grad V)}{l\Gamma}\right) \right]D_l(\vec k) = 0,
\end{equation}
where $\varepsilon_{q\Gamma (\vec k)}$ is the energy of the q$^{th}$ band at $\Gamma$ point ($\vec k$ point). The sum runs over the $|l\Gamma\rangle$ Bloch states at $\Gamma$, where $D_l(\vec k)$ is the wave vector-dependent expansion coefficient. The term with $(\vec \sigma \times \grad V)$ accounts for the spin-orbit coupling, where $V$ determines the crystalline periodic potential. For a basis formed by the linear combination of four band edge Bloch functions~\cite{bastard1990wave}, we can write eq.~\ref{eq:KP_Hamiltonian} as a hermitian matrix~\cite{bastos2016stability}

\begin{equation}\label{eq:KP_Hamiltonian_matrix}
    H\left(\vec k\right) = \begin{bmatrix}
T & -S& i\frac{T-Q}{\sqrt{2}} & -i\sqrt{\frac{2}{3}}P_z\\
 & Q & -i\frac{S^\dagger}{\sqrt{2}}& -P_+ \\
 &  & \frac{Q+T}{2} + \Delta & -\frac{P_z}{3} \\
 &  &  & E_\Gamma 
\end{bmatrix}, 
\end{equation}
where 

\begin{eqnarray}
T = -\frac{\hbar^2}{2m^*}(\gamma_1-\gamma_2)(k_x^2+k_y^2) + (\gamma_1+2\gamma_2)k_z^2\nonumber,\\
S = i\frac{\hbar^2}{2m^*}(2\sqrt{3}\gamma_3  k_z  k_-),\nonumber\\
Q = \frac{\hbar^2}{2m^*}(\gamma_1+\gamma_2)(k_x^2+k_y^2) + (\gamma_1-2\gamma_2)k_z^2,\\
P_z = Pk_z,\nonumber\\
E_\Gamma = \epsilon_\Gamma + \frac{\hbar^2}{2}(k_x^2+k_y^2+k_z^2)\left\{ \frac{1}{m^*} - \frac{E_p}{3}\left[\frac{2}{\epsilon_\Gamma}  + \frac{1}{\epsilon_\Gamma + \Delta}\right] \right\},\nonumber
\end{eqnarray}
where $\gamma_1$, $\gamma_2$, and $\gamma_3$ are empirical Luttinger parameters,  $\varepsilon_\Gamma$ is the band gap at the Brillouin zone center, $P$ is related to the Kane energy $E_p$, $k_\pm=(k_x \pm i k_y)/\sqrt{2}$, $\Delta$ is the spin-orbit coupling of the valence band, and $m^*$ is the effective mass. The diagonalization of eq.~\ref{eq:KP_Hamiltonian_matrix} for wave vectors within the Brillouin zone returns the electronic band dispersion. For compounds of III-V family, the $k\cdot p$ method has been shown to return accurate band structures around high symmetry points~\cite{bastos2016stability}.

\subsection{Subspace-Search Variational Quantum Eigensolver}

\begin{figure}[!htb]
    \centering
    \label{SSH_ANSATZ3}\begin{tikzpicture}
        \node[scale=1.0] {
        \begin{quantikz}
            \lstick[wires=2]{$\ket{\psi}^{\otimes n}$} & \gate{U(\theta_1, \theta_2, \theta_3)} \gategroup[2,steps=3,style={dashed,rounded corners,fill=green!20, inner xsep=2pt},background,label style={label position=below,anchor=north,yshift=-0.2cm}]{{\sc 
            N Layers}} & \targ{} & \ctrl {1} &  \meter{} \gategroup[2,steps=1,style={dashed,rounded corners,fill=blue!15, inner xsep=2pt},background,label style={label position=below,anchor=north,yshift=-0.2cm}]{{\sc qwc}}   \\
            & \gate{U(\theta_4, \theta_5, \theta_6)} & \ctrl{-1} & \targ & \qw & \meter{} 
        \end{quantikz}
        };
    \end{tikzpicture}\\
    \caption{(color online) Schematics of the circuit employed to obtain the eigenenergies of the Hamiltonian. The left green box highlights the strongly entangled states used as the ansatz composed of N layers and n qubits. The right blue box highlights the measurement scheme using qubit-wise commutativity (QWC).}
    \label{fig:ansatz}
\end{figure}
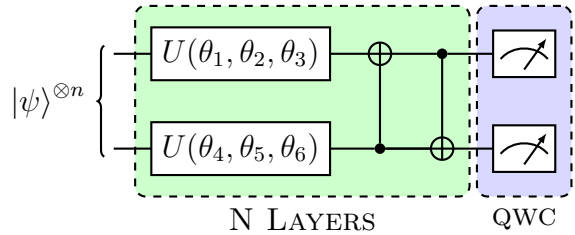

The Variational Quantum Eigensolver (VQE) method has been shown to be a powerful tool to obtain the ground state of Hamiltonians using noisy intermediate-scale quantum (NISQ) computers~\cite{peruzzo2014variational}. It is an hybrid quantum-classical algorithm where the expectation value of the Hamiltonian under analysis, $\expval{H(\vec k)},$ is minimized using together a quantum and a classical computer.
For a parameterized ansatz circuit $U(\vec\theta)$ (fig.\ref{fig:ansatz}), the quantum computer generates the ansatz state $\ket{\phi(\vec\theta)}=U(\vec\theta)\ket{\psi}$ such that the expectation value $\expval{H(\vec k, \vec\theta)}=  \expval{H(\vec k)}{\phi(\vec\theta)}$ is minimized iteratively, feeding $U(\vec\theta)$ back with optimization parameters $\vec\theta$ gathered from the classical machine. The minimized expectation value of the Hamiltonian tends to the ground-state energy of the system. 

However, to determine the band structures, not only the ground state is desirable, but also the excited states. To assess both the ground and the excited states of the Hamiltonian of Eq.~\ref{eq:KP_Hamiltonian_matrix} utilizing a quantum computer, we employ the Subspace-Search Variational Quantum Eigensolver (SSVQE)~\cite{nakanishi2019subspace}. The SSVQE method refines the VQE method to obtain excited states without substantially raising computational cost, which is advantageous with NISQ machines.

The motivation behind the SSVQE method is to perform a single optimization procedure, and use the optimized parameters $\vec\theta$ to evaluate the expectation value of the Hamiltonian for a group of initially orthonormal states. First, for $n$ qubits we employ as the ansatz $U(\vec\theta)$ a strongly entangled circuit~\cite{schuld2020circuit}, consisting of N layers of single qubit rotations and entanglers (fig.~\ref{fig:ansatz}). Then, p-copies of the circuit are initialized with p-orthogonal states $\ket{\psi_p}=\ket{\psi}^{\otimes n}$, with p representing the number of excited states to be evaluated. A weighted cost function $C_w=\sum_l^p w_l \expval{U^{\dagger}(\vec\theta) H U(\vec\theta)}{\psi_l}$ is minimized with the weights $w \in [l, l-1,\cdots,1]$. Once the minimization attains to a global minimum, the ansatz becomes unitary mapping the pre-orthonormal states $\ket{\psi_p}$ to the orthonormal eigenstates of the Hamiltonian.

To obtain the expectation values used with the cost function, the Hamiltonian of Eq.~\ref{eq:KP_Hamiltonian_matrix} is decomposed using the Jordan-Wigner decomposition~\cite{nielsen2005fermionic} in terms of a linear combination of Pauli matrices, $H=\sum_j h_j \sigma^{\otimes n}$. We apply qubit-wise commutativity (QWC) to reduce the number of measurements~\cite{izmaylov2019unitary}.

\section{RESULTS}

\subsection{Band Structures}

\begin{figure}
    \centering
    \includegraphics[scale=0.4]{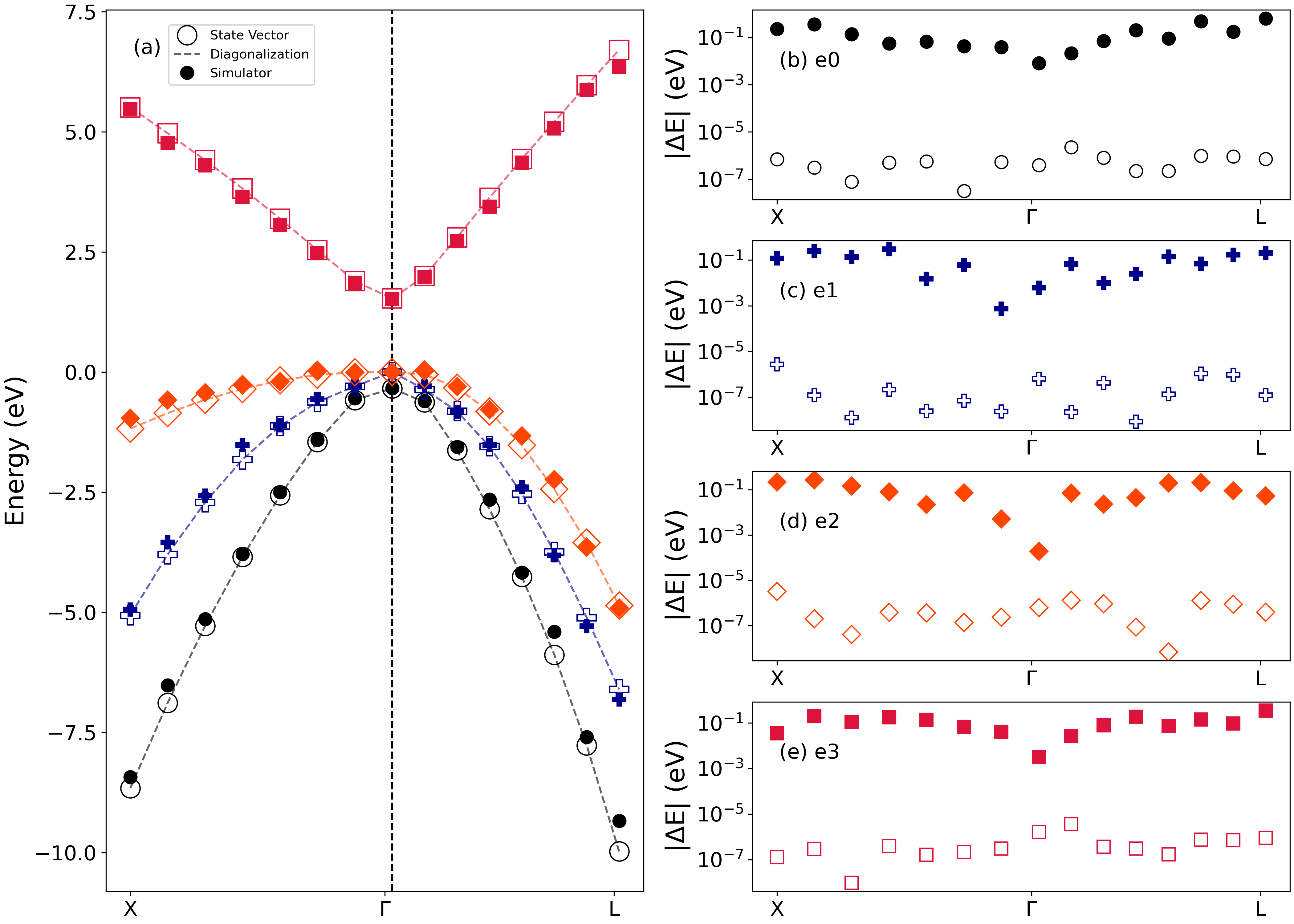}
    \caption{(a) GaAs band structure for the lower conduction band state (red squares), and top heavy hole (blue pluses), light hole (orange diamonds) and split-off (black circles) valence band states, evaluated using both the state vector solver (open symbols), the probabilistic simulator (full symbols), and numerical diagonalization (dashed lines). Absolute energy difference between the results from the numerical diagonalization and the state vector solver (open symbols), and from the numerical diagonalization and from the probabilistic simulator (full symbols) for (b) the split-off band, (c) the heavy hole band, (d) the light hole, and (e) the conduction band.}
    \label{fig:GaAs_4layers_with_error}
\end{figure}

Here we utilized the PennyLane framework~\cite{bergholm2018pennylane} to create and execute the quantum circuits. Figure~\ref{fig:GaAs_4layers_with_error}(a) shows the band structure obtained for the Gallium Arsenide (GaAs) compound using the state vector solver (open symbols), the probabilistic simulator (full symbols), and the direct diagonalization (solid lines). The structural parameters used with the Hamiltonian of Eq.~\ref{eq:KP_Hamiltonian_matrix} are found in the supplementary table S1. Employing an ansatz with 5 layers (see Fig.~\ref{fig:ansatz}), the band structure is evaluated for a uniformly distributed path across the first Brillouin zone through the high-symmetry points $X-\Gamma-L$, for the split-off band (black curve), the heavy hole band (blue curve), light hole band (orange curve), and conduction band (red curve). For the classical optimization process, we use the Adam optimizer with an step rate 0.01. We stop the optimization process when a difference of 10$^{-7}$ eV among two consecutive steps is reached. The agreement between the results from the SSVQE and the direct diagonalization proved to be excellent.  Figs.~\ref{fig:GaAs_4layers_with_error}(b-e) show the absolute energy differences between the quantum simulation, using both the state vector solver (open symbols) and the probabilistic simulator (full symbols), with respect to the expected eigenvalues for each band (dashed lines). The state vector solver represents the best scenario for the simulation since it returns the exact solution expected for the quantum system. For it, the maximum energy difference ranges for up to tens of mili-electron volts for the worst case of the light hole band next to the $X$ point (Fig.~\ref{fig:GaAs_4layers_with_error}(d)).  

To acquire a better picture of the expected results of a quantum device, we calculated the GaAs band structure using the probabilistic simulator~\cite{bergholm2018pennylane} (full symbols in fig.~\ref{fig:GaAs_4layers_with_error}(a)), that samples the evaluation of the expectation values for a finite number of shots, retrieving the probabilistic nature of the quantum measurement. We also adopt the Adam optimizer for an ansatz with 5 layers. The use of the probabilistic simulator bring difficulties for the optimization procedure. The simulation is unable to converge with the same precision of the state vector solver, since the probabilistic fluctuation inherent to quantum processes is added to the measurement results. Therefore, we lowered the convergence threshold to $1e^{-4}$ eV, and use 10000 shots. Furthermore, the simulation time increases considerably. The mean time of each cycle raised to 3480 s, while the mean number of cycles is 97. As we can observe in fig.~\ref{fig:GaAs_4layers_with_error}(a), the results from the probabilistic simulator (full symbols) do not perfectly match the ones form the exact diagonalization (dashed lines). The error of the probabilistic simulator in comparison to the state vector solver results  is increased, with a maximum of 400 meV for the light hole band around the X high symmetry point (fig.~\ref{fig:GaAs_4layers_with_error}(d)).

Once we demonstrate the possibility of obtaining the band structure of bulk crystalline semiconductor structures using a quantum circuit, in what follows, we present a time benchmark for the use of the SSVQE method with k$\cdot$p crystalline bulk Hamiltonians.

\subsection{Execution Times}

As previously advocated, the iterative hybrid quantum-classical SSVQE method relies on minimizing the expectation value for a parameterized quantum circuit, carrying a search in the parameter space using classical optimizers. The parameter space size, defined by the ansatz depth, bounds the performance of the minimization process. For each point in the k-path, the execution time is a trade between the time invested running the quantum circuit to recover the expectation value, and the number of optimization cycles necessary to minimize it. Shallow ansatzes require a smaller parameter space, facilitating the optimization procedure. However, owing to the small amount of adjustable parameters, the optimizer might need additional optimization cycles, which raises the number of measurements and also the computation time within the quantum machine. In turn, deeper ansatzes may be more generic, allowing for fewer optimization steps yet demanding costly expectation value evaluations. Choosing the ideal scenario depends upon the available resources. For NISQ computers, the hardware restricts the circuit depth since decoherence times should be greater than the time spent performing the circuit transpiling and execution ~\cite{preskill2018}. Additionally, up to now the access to quantum computers is mainly restricted to cloud services, with job-based queues. After each expectation values for each cycle of the optimization, the job is sent back to the queue. Therefore, minimizing both the number of measurements in the expansion of the Hamiltonian, and the number of cycles, while keeping the ansatz shallow, is mandatory.

\begin{figure}[htb!]
    \centering
    \includegraphics[scale=0.4]{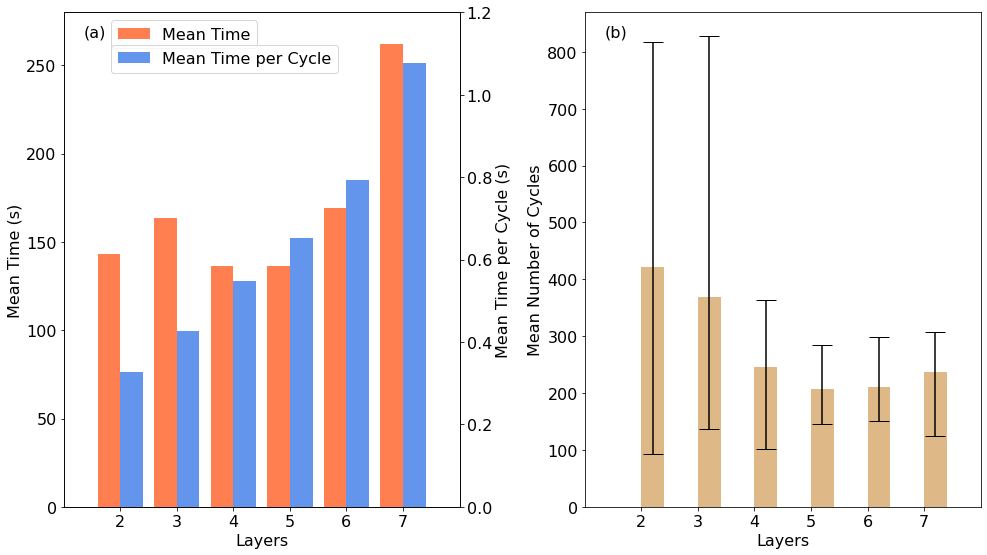}
    \caption{(a) In orange, mean time in seconds for convergence of each point in the k-path for 2-7 layers in the quantum circuit, and in blue the mean time to complete each cycle of convergence attempt. (b) In the bars we have the mean number of cycles to complete the convergence, for all the k-path for 2-7 layers in the quantum circuit, the black bars represent the maximum and minimum number of cycle for a point.}
    \label{fig:Optimization Times}
\end{figure}

Figure~\ref{fig:Optimization Times} shows the mean times necessary to obtain the band structure of the GaAs bulk Hamiltonian, in terms of the number of layers of the ansatz and using the state vector solver. The mean is done taking into account all points in the k-path (see Fig.~\ref{fig:GaAs_4layers_with_error}).  Overall, the mean execution time remains around 150 s for ansatz with up to 5 layers, escalating significantly to 262 s for 7 layers (Fig.~\ref{fig:Optimization Times}(a)). As expected, the mean time per cycle rises monotonically with the number of layers, since the number of gates employed in the quantum circuit also increases. Figure~\ref{fig:Optimization Times}(b) shows the mean number of cycles with respect to the number of layers. The horizontal lines in each bar represent the minimum and maximum number of cycles. For 2 layers, a shallow ansatz demands the greater mean number of cycles (420), with a maximum of 818 cycles for one specific k-point in the path. For 5 layers, we see the lower mean number of cycles (207) with a maximum of 285 cycles and a minimum of 145 cycles. The number of cycles, maximums and minimums rise back for 6 and 7 layers. As we can observe, the strongly entangled ansatz with 5 layers reveals to be the more appropriate for simulating 4$\times$4 k$\cdot$p Hamiltonians.    

The choice of the optimizer is also fundamental, since its performance determines the number of circuit evaluations. Supplementary figure S1 shows the mean execution times for the optimizers: Adam, Adagrad, Conjugated Gradient, and Nesterov. The optimization is done using 5 layers and 0.01 for the step rate, with a tolerance of 10$^{-7}$ eV. The mean time per cycle remains constant since it depends on expectation value evaluation using the quantum circuit. However, the Adam optimizer shows the lowest mean time (137 s), while the Conjugated Gradient shows the bigger one (1134 s) (Fig S1(a)). The short mean time obtained by the Adam optimizer is related to its efficiency in finding the total minimum with the small number of steps (248 on average). Since the band structure is evaluated from the eigenvalues of Hamiltonians that differ only from each other by the neighbor k-points, it is worth mentioning that we used the optimized parameters from the previous k-point as a starting point for optimizing the next k-point.  

\begin{figure}[!h]
    \centering
    \includegraphics[scale=0.35]{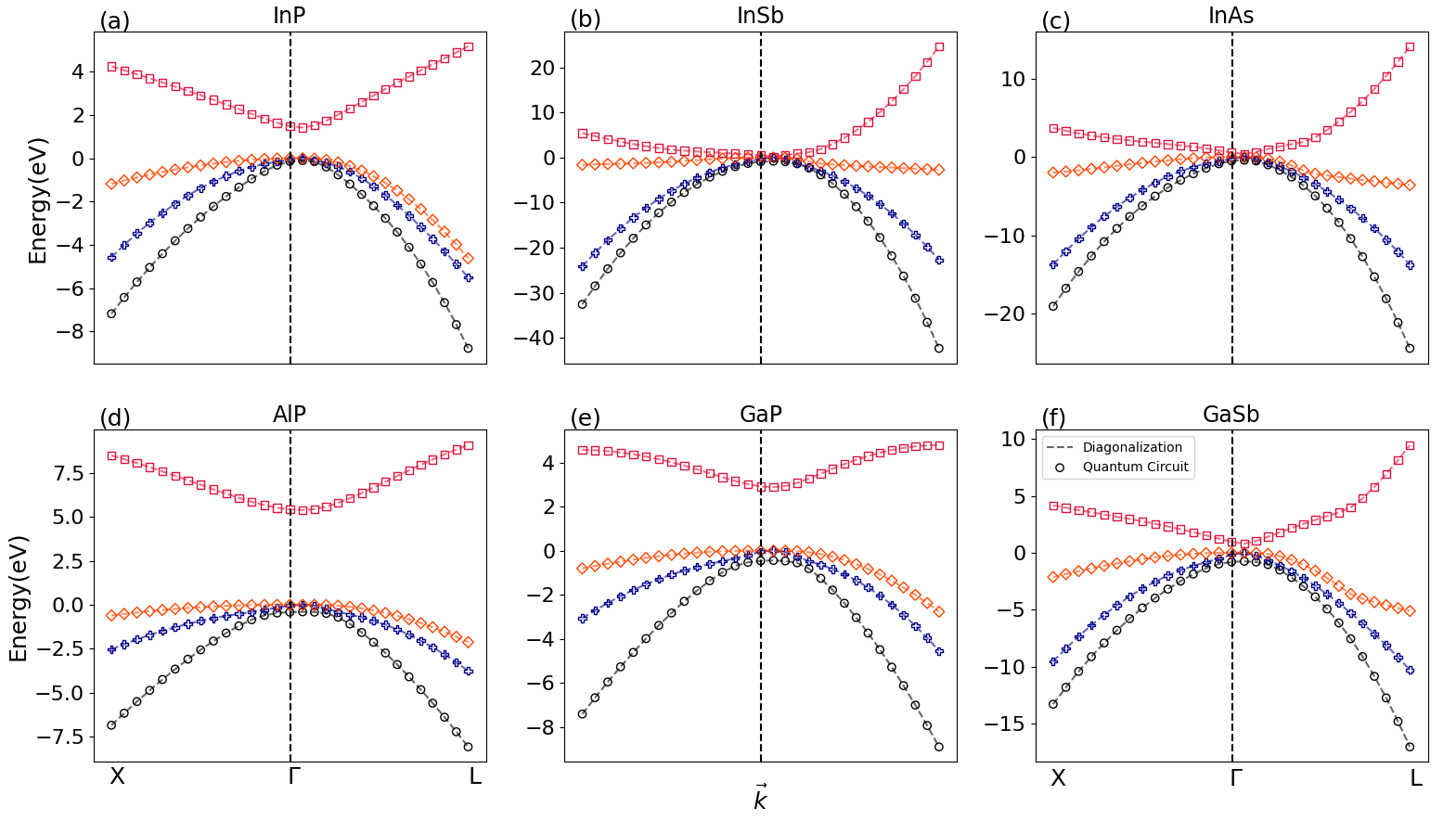}
    \caption{Band structure for different materials evaluated using both the quantum circuit (dots) and numerical diagonalization (solid lines). On (a) we have the result for InP structure, (b) for InAs, (c) InSb, (d) AIP, (e) GaP and (f) for GaSb. All the structures were evaluated using the Penny Lane State Vector.}
    \label{fig:all_structures}
\end{figure}

After validating the application of the SSVQE method to obtain the band structure of the GaAs structure, we applied the Adam optimizer and the strongly entangled ansatz with 5 layers to evaluate the electronic structure of other compounds of the III-V family. Figure~\ref{fig:all_structures} shows the band structure for the Indium Phosphide (InP, fig.~\ref{fig:all_structures}(a)), the Indium Arsenide (InAs, fig.~\ref{fig:all_structures}(b)), the Indium Antimonide (InSb, fig.~\ref{fig:all_structures}(c)), the Aluminium Phosphide (AlP, fig.~\ref{fig:all_structures}(d)), the Galium Phosphide (GaP, fig.~\ref{fig:all_structures}(e)), and the Galium Antimonide (GaSb, fig.~\ref{fig:all_structures}(f)). For all graphs in fig.~\ref{fig:all_structures} the dashed lines are the results from the direct diagonalization of the Hamiltonian employing the specific parameters for each compound, and the symbols are the results for the quantum circuit simulation. Once, the agreement between the expected and the simulated values is excellent. Supplementary figure S2, shows the energy differences. As for the GaAs structure, the error is small, with a maximum value of 4 meV for the heavy hole band of the InP compound (see supplementary figure S2(a).

\subsection{Quantum Device}

Besides the ability of obtaining the band structures using both the state vector and the probabilistic simulators, we also used a real-device-based noise model to simulate the expected response of a real quantum computer. Namely, we exploit the \textit{Qiskit Aer noise} module to generate a noise model as an approximation of the real errors in the execution of the \textit{ibmq-lima} quantum device~\cite{IBMq}. The \textit{ibmq-lima} is an IBM's quantum processor of Falcon type, with 5 qubits, and  94.03$\mu$s (101.39$\mu$s) of average T1 (T2) error. 
Supplementary figure S3 shows the results simulated using the \textit{ibmq-lima} noise model. As in the case of the probabilistic simulator, the convergence of SSVQE method is hampered by the fluctuations of the quantum measurement. The convergence threshold is lowered to $10^{-3}$~eV with 10000 shots, but the model marginally attaches to a global minima. The energy differences between the device and the exact diagonalization are mainly in the range of tens of electron volts. Moreover, the worst scenario is for the electron band (the more energetic), with an error of 1.8~eV for the high-symmetry point $L$ (see supplementary figure S2(e)). Each cycle of the simulation using the noise model is time consuming. The execution time increased considerably in comparison to the probabilistic simulator, taking around 12h to simulate each point within the k-path.

\subsection{Absorption}

\begin{figure}
    \centering
    \includegraphics[scale=0.4]{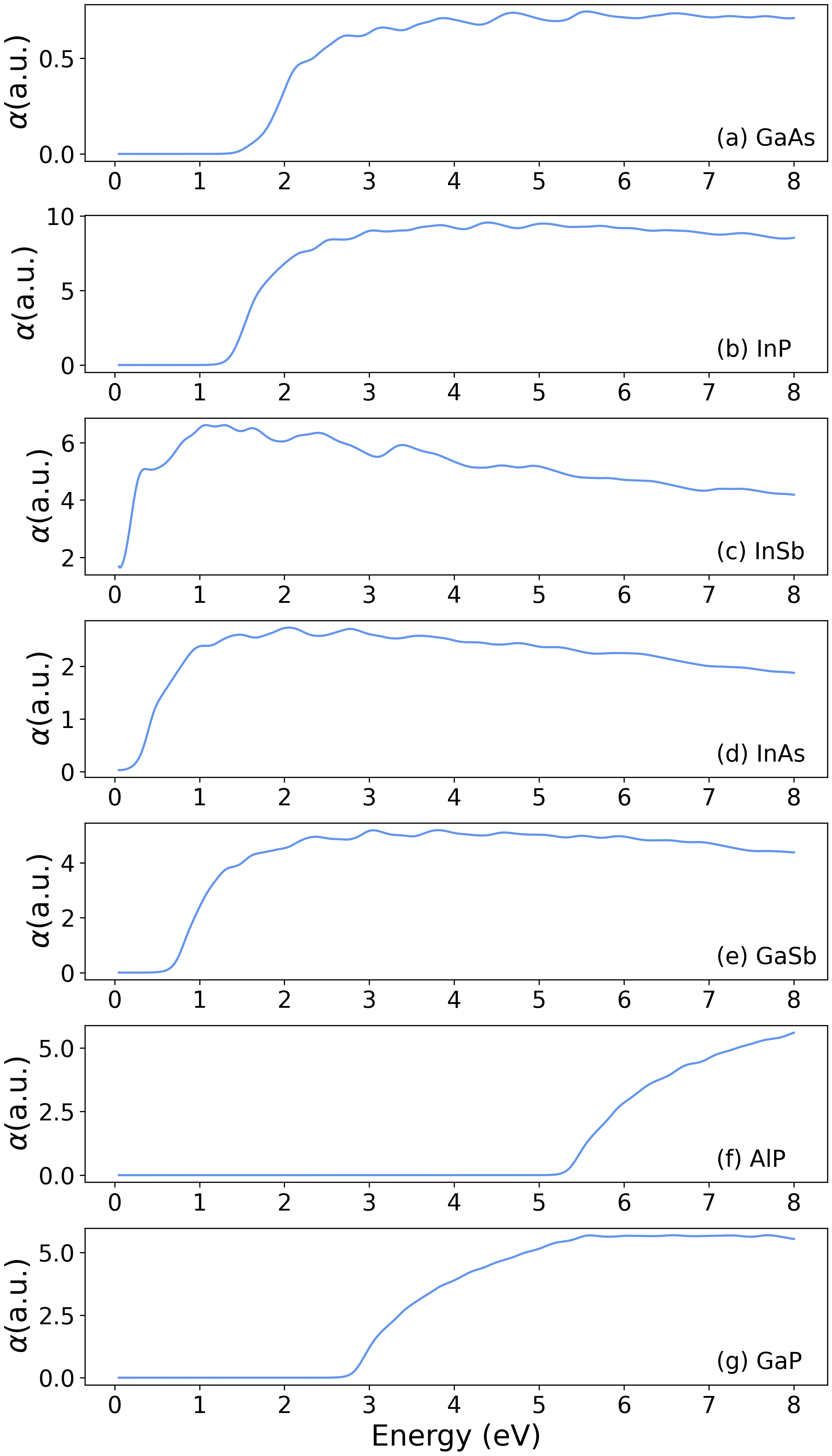}
    \caption{Simulated interband absorption of semiconductor compounds (a) GaAs, (b) InP, (c) InSb, (d) InAs, (e) GaSb, (f) AlP, and (g) GaP.}
    \label{fig:absorption}
\end{figure}

NISQ computers, aided by variational methods as the SSVQE, are presumed to generate ground and excited state wave functions not classically feasible~\cite{ibe2022calculating}. With them, transition amplitudes, related to experimental properties as emission and absorption, can be readily evaluated. Noting that $\expval{\hat T_{vc}} = \mel{\psi_v}{\hat T}{\psi_c} = \mel{\phi_v}{U^{\dagger}(\vec\theta)\hat T U(\vec\theta)}{\phi_c}$ determines the expected value for the transition operator $\hat T$ between valence $\ket{\psi_v}$ and conduction $\ket{\psi_c}$ states, where $U(\vec\theta)$ represents the optimized ansatz, the evaluation of the transition amplitude can be written as~\cite{nakanishi2019subspace}

\begin{equation}
\begin{split}
    \Re{\mel{\phi_v}{\Tilde{T}(\vec\theta) }{\phi_c}}&=\\
    &\mel{\phi_{+,vc}^{x}}{\Tilde{T}(\vec\theta)}{\phi_{+,vc}^{x}}-\frac{1}{4}\mel{\phi_v}{\Tilde{T}(\vec\theta) }{\phi_v}\mel{\phi_c}{\Tilde{T}(\vec\theta) }{\phi_c},
\end{split}
\end{equation}
\begin{equation}
 \begin{split}
    \Im{\mel{\phi_v}{\Tilde{T}(\vec\theta) }{\phi_c} }&=\\
    &\mel{\phi_{+,vc}^{y}}{\Tilde{T}(\vec\theta)}{\phi_{+,vc}^{y}}-\frac{1}{2}\mel{\phi_v}{\Tilde{T}(\vec\theta) }{\phi_v}-\frac{1}{2}\mel{\phi_c}{\Tilde{T}(\vec\theta) }{\phi_c},
\end{split}   
\end{equation}
where $\Tilde{T}(\vec\theta)=U^{\dagger}(\vec\theta)\hat T U(\vec\theta)$, $\sqrt 2\ket{\phi_{+,vc}^{x}} = \ket{\phi_v}+\ket{\phi_c}$, and $\sqrt 2\ket{\phi_{+,vc}^{y}} = \ket{\phi_v}+i\ket{\phi_c}$. Note that obtaining the transition rate involves few additional evaluations of the quantum circuit for the optimized set of parameters $\vec\theta$. To evaluate the absorption, in the case of interband transitions, the transition amplitude (or the oscillator strength) is given by the overlap between the initial and final states. Therefore, we can assume the transition operator $\hat T$ as the identity operator. 

The absorption reads~\cite{bezerra2017lifetime}

\begin{equation}
    \alpha(\hbar\omega) \propto \frac{1}{\hbar\omega}\sum_{v,\vec k} \abs{\mel{\phi_v(\vec k)}{\Tilde{T}(\theta) }{\phi_c(\vec k)}}^2\Theta(\varepsilon_{vc}(\vec k)-\hbar\omega),
\end{equation}
where the sum runs over the valence band states, and over the $\vec k$ point within the path. $\Theta(\varepsilon_{vc}(\vec k)-\hbar\omega)$ is a step function with $\varepsilon_{vc}(\vec k) = \varepsilon_{c} - \varepsilon_{v}$ being the energy separation between the valence and the conduction band states at $\vec k$ point. Figures~\ref{fig:absorption}(a)-(g) show the interband absorption between the valence and conduction bands for several semiconductor compounds, simulated using the wave functions from the quantum simulation of the band structures. Even considering the contribution of a small amount of bands, the results are in good agreement with the expected absorption spectrum for semiconductor compounds of the III-V family. Simulating the absorption using a quantum computer, for an optimized ansatz, is a considerably cheap operation in terms of system requirements. For each point in k-path, to evaluate the absorption is necessary only 10 calls of the quantum computer, barely the same number of calls used in one cycle of the optimization process.

\section{DISCUSSION}

We show SSVQE method as a viable, yet costly, tool for obtaining the band structure of semiconductor bulk compounds using a quantum computer. Despite it being an algorithm specially designed to operate on NISQ computers, we need to be careful when dealing with band structures of crystalline materials. As we can observe, using the k$\cdot$p method, we are able to deal with small Hamiltonians, which is advantageous with the limited computational power of NISQ processors~\cite{daley2022practical}, still capturing relevant information about electronic movement within the material around high symmetry points. Notwithstanding, even for such small Hamiltonians, the simulation times are not competitive with traditional diagonalization methods. 

The increased simulation time is related to both the number of cycles in the optimization process, and the number of terms of the Hamiltonian decomposed as a combination of Pauli strings~\cite{ibe2022calculating}. For each step of the optimization process and for each term in the decomposition, an evaluation of the quantum circuit is required. To reduce the number of cycles we should play with the optimization process itself. We have to choose the appropriated optimizer, the appropriated precision for the convergence, and, primarily, the appropriated ansatz circuit. The number of qubits raises with the complexity of the Hamiltonian, increasing considerably the number of decomposition terms. To reduce the number of terms, quantum-wise commuting (QWC) scheme can be used. It employs family commutativity to group the terms required to obtain the expectation values ~\cite{gokhale2019minimizing}. The QWC, on average, reduces a large number of terms by a factor of three in comparison to the total number of Pauli strings. Since we are dealing with only two qubits, here we reduced the number of terms from 16 to 9. More efficient methods can be applied, such as the general commutativity (GC), which tends to reduce the number of terms by a factor of ten for large Hamiltonians~\cite{gokhale2019minimizing}. However, such methods use graphs and coloring theory which are themselves NP-hard problems,  potentially increasing the time in the classical step of the simulation~\cite{verteletskyi2020measurement}. Fortunately, quantum computers can also aid with a significantly speedup on NP-hard problems using, for example, Grover's algorithm~\cite{anikeeva2021number}.

When moving from the state vector solver to a probabilistic simulator or to a device simulator, the difficulties increase considerably. The parameters' optimization landscapes for VQE-based methods are non-convex. Therefore, reaching a minimum requires at least hundreds of steps. For the calculation of excited state with the SSVQE method, the weighted cost function further increases the optimization challenge~\cite{lyu2022symmetry}, in special with noisy devices. Since the cost relies on the summation of noisy expectation values~\cite{nakanishi2019subspace}, the variance in the total energy gains a stochastic component hampering the optimization routine. An additional problem is related to the evaluation of gradients of the cost function that tends to vanish, prejudicing the action of gradient based optimizers. Such phenomenon is referred to as barren plateaus~\cite{cerezo2021cost,wang2021noise}. A possible detour would be to employ gradient-free methods for the optimization. Nevertheless, even gradient-free methods have been shown to present barren plateaus~\cite{arrasmith2021effect}.      

As we can observe from our results, we are able to accurately obtain the band structure of semiconductor bulk compounds using the SSVQE method. However, this is not an easy task to be implemented on a real NISQ device, at least in a competitive fashion. There are several aspects to consider, emphasizing the importance of the design of structure- and method-dependent approaches~\cite{sherbert2021systematic,sherbert2022orthogonal}. The evaluation of excited states imposes additional difficulties with more complex cost functions. Fortunately, incorporating symmetries of the Hamiltonian in the cost function proved beneficial~\cite{lyu2022symmetry}. Nevertheless, barren plateaus represent a challenging to be surpassed. They have been show to be a problem also with hardware efficient ansatz~\cite{mcclean2018barren}. Beyond the plateaus originated by the non-convex parameters' landscape, noise-induced plateaus have been shown to appear with NISQ devices~\cite{wang2021noise}.  
The optimization process itself deserves a special care. There are some proposals of more robust techniques to be employed with NISQ devices. The Quantum Natural Gradient (QNG) for example, tends to reduce the number of optimization steps in comparison to gradient-based methods by computing the Fubiny-Study metric tensor~\cite{stokes2020quantum}. However, each step of the optimization requires more calls from the quantum computer, which can be a problem for NISQ computers. Alternatively, in the Quantum Analytic Descent approach~\cite{koczor2022quantum} the optimization landscape can be approximated in the vicinity of reference points, allowing for the cheaper evaluation of the gradients using the classical computer. The quantum device is exclusively employed to evaluate the cost function at the reference point and for some neighbor points, being the Quantum Analytic Descent method closer to a classical than to a pure quantum algorithm. Just as importantly, the stochastic nature of the measurement is further explored in the Doubly Stochastic Gradient Decent method~\cite{sweke2020stochastic}. As a consequence, fewer measurements and circuits executions are required yet ensuring convergence for restricted settings. However, the effect of noise is not included in the Doubly Stochastic Gradient Decent method, which can be an impediment with NISQ computers.  

Besides the difficulties, the NISQ computers' architecture can be exploited in favor of the band structures of bulk crystalline systems, specially when utilizing k$\cdot$p-based methods. NISQ computers are designed with a reduced quantity of qubits forming a not entirely connected grid. As mentioned earlier, k$\cdot$p Hamiltonians are in general small, still preserving important aspects of the electronic structure of crystalline materials. Therefore, a reduced number of qubits is sufficient to simulate the Hamiltonians. For instance, a 4$\times$4 Hamiltonian demands only 2 qubits. To acquire quantitatively correctly description of valence and conduction bands, slightly bigger Hamiltonians are needed. For example, a precise description of valence bands and two lowest conduction bands of direct and indirect gap bulk semiconductors can be obtained with 24$\times$24 basis~\cite{ben2003band}. Nonetheless, a  40$\times$40 band k$\cdot$p is enough to accurately reproduce the overall band structure of strained and unstrained semiconductor alloys~\cite{neffati2012full,gawarecki2022invariant}. Even a big 40$\times$40 k$\cdot$p Hamiltonian, from the quantum computer perspective, requires solely 6 qubits.  

Moreover, since one challenge with hybrid algorithms is to reduce the number of calls of the quantum device, a crude parallelization approach can be used. Considering the decomposition of the Hamiltonian as a combination of non-commuting Pauli strings, groups of measurements can be done simultaneously using untangled groups of qubits of the same chip, taking advantage of single calls of the quantum device. Here, the absence of connections among the qubits is not a complication, provided that the measurements should be done individually for non-commuting observables.

\section{CONCLUSION}

In summary, we presented the use of NISQ computers to obtain the band structures of crystalline semiconductors of the III-V family. The structures are described in terms of k$\cdot$p Hamiltonians, whose expectation values are minimized using the SSVQE method. The execution times are closely dependent on the optimization parameters, the depth of the ansatz, the number of terms of the Hamiltonian decomposition, and the quantum simulator employed. Using the state-vector solver, we observe an excellent agreement between the simulated band structures and the expected ones. For the probabilistic simulator and the simulator of a real NISQ computer, the minimization is intricate, considerably raising the overall simulation time and lowering the precision, still returning good results. The use of k$\cdot$p Hamiltonians is advantageous with NISQ devices. It allows for a precise description of the electronic properties of semiconductor structures using a small number of qubits. Furthermore, with the optimized ansatzes, we obtain the interband absorption spectra of the structures. It requires only a few additional runs of the quantum device, demonstrating the possibility of using quantum computers to readily obtain the properties of crystalline systems. Using k$\cdot$p Hamiltonians opens up the possibility of using NISQ computers to study the influence of strain, stress, and confinement in the semiconductor structures~\cite{gladysiewicz20158}.

\section{CODE AVAILABILITY}

The code is available under request to the corresponding author.

\section*{ACKNOWLEDGMENTS}
We acknowledge the use of IBM Quantum services for this work. The views expressed are those of the authors, and do not reflect the official policy or position of IBM or the IBM Quantum team. Our special thanks to the Coordena\c{c}\~ao de Aperfei\c{c}oamento de Pessoal de N\'ivel Superior (CAPES) and to the Funda\c{c}\~ao de Amparo \`a Pesquisa e Inova\c{c}\~ao do Estado de Santa Catarina (FAPESC) for providing resources to this research. 

\bibliographystyle{plain}
\newcommand{\newblock}{}
\bibliography{refs.bib}

\end{document}


\title[Supplementary Material: Revisiting semiconductor bulk hamiltonians using quantum computers]{Supplementary Material:  Revisiting semiconductor bulk hamiltonians using quantum computers}

\author{Raphael César de Souza Pimenta$^1$ \& Anibal Thiago Bezerra$^2$}

\address{$^1$ Departamento de Física, Universidade Federal de Santa Catarina,  Roberto Sampaio Gonzaga s/n, Florianópolis-SC, Brasil}

\address{$^2$ Departamento de Física, Universidade Federal de Alfenas, Jovino Fernandes Sales 2600, Alfenas-MG, Brazil}

\ead{anibal.bezerra@unifal-mg.edu.br}
\vspace{10pt}
\begin{indented}
\item[]April 2022
\end{indented}

\vspace{2cm}
\begin{table}[htb!]
\centering
\caption{Structural parameters for III-V compounds from ref.~\cite{bastard1990wave,bastos2016stability,sytnyk2018luttinger}}
\label{tab:III-V_parameters}
\begin{tabular}{@{}llllllll@{}}
\toprule
                     & GaAs   & GaSb   & InP    & InAs  & InSb   & AlP & GaP\\ \midrule
$\varepsilon_0$ (eV) & 1.5192 & 0.811  & 1.4236 & 0.418 & 0.2352 & 5.4 & 2.88\\
$\Delta$ (eV)        & 0.341  & 0.752  & 0.108  & 0.38  & 0.81   & 0.4 & 0.45\\
$m^*$                & 0.0665 & 0.0405 & 0.079  & 0.023 & 0.0139 & 0.7 & 0.79\\
$E_p$ (eV)           & 22.71  & 22.88  & 17     & 21.11 & 22.49  & 27.1 & 22.2\\ 
$\gamma_1$           & 7.03   & 13.4   & 6.64   & 19.81 & 34.8   & 3.31 & 4.02\\
$\gamma_2$           & 2.33   & 4.7    & 2.1    & 8.46  & 15.66  & 0.71 & 0.98\\
$\gamma_3$           & 3.03   & 6      & 2.81   & 9.34  & 16.58  & 1.23 & 1.66\\\bottomrule
\end{tabular}
\end{table}

\begin{figure}[htb!]
    \centering
    \includegraphics[scale=0.4]{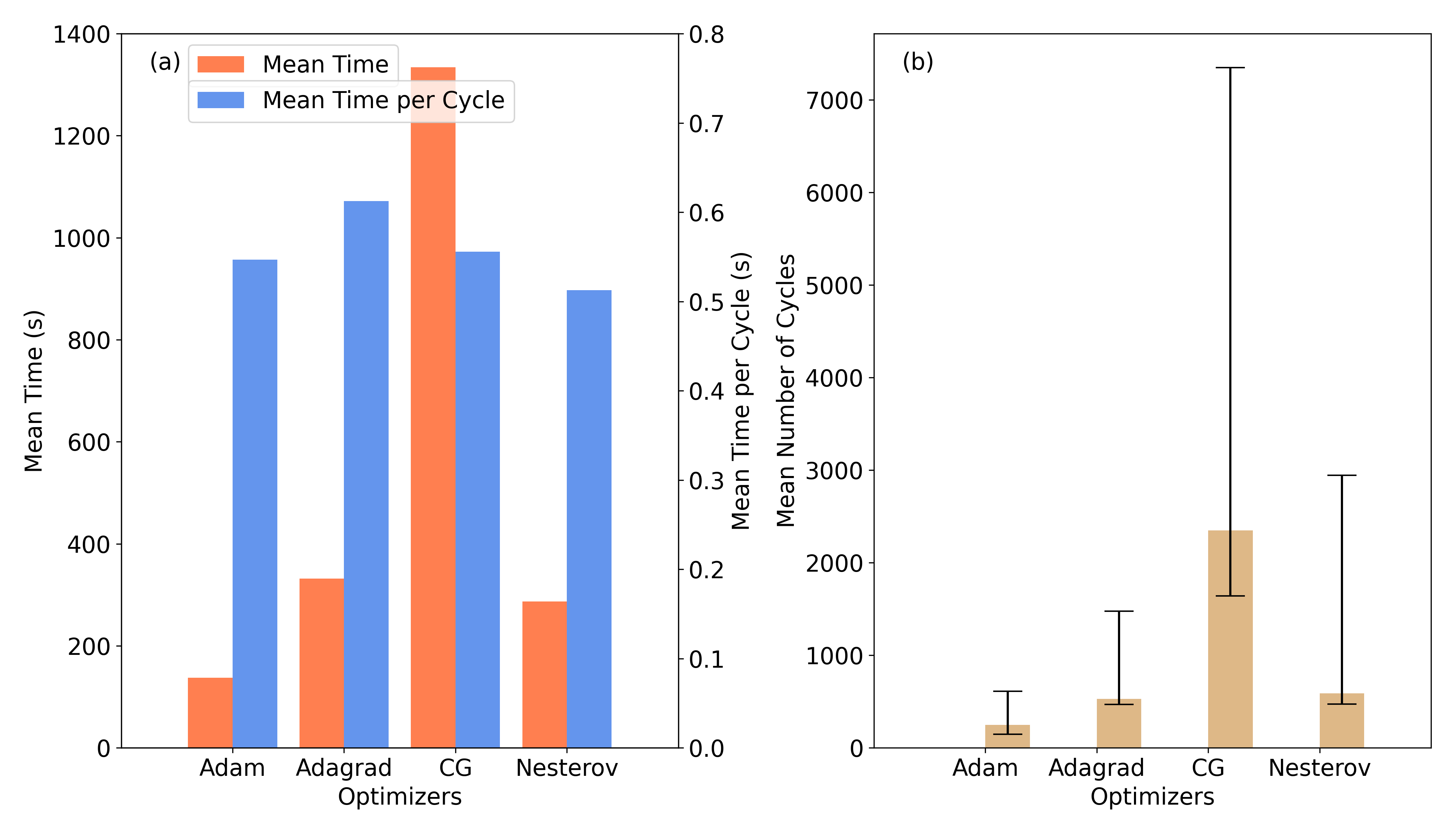}
    \caption{(a) Mean simulation time and mean time per cycle for different optimizers. (b) Mean number of cycles for different optimizers.}
    \label{fig:suppleTimes}
\end{figure}

\begin{figure}[htb!]
    \centering
    \includegraphics[scale=0.4]{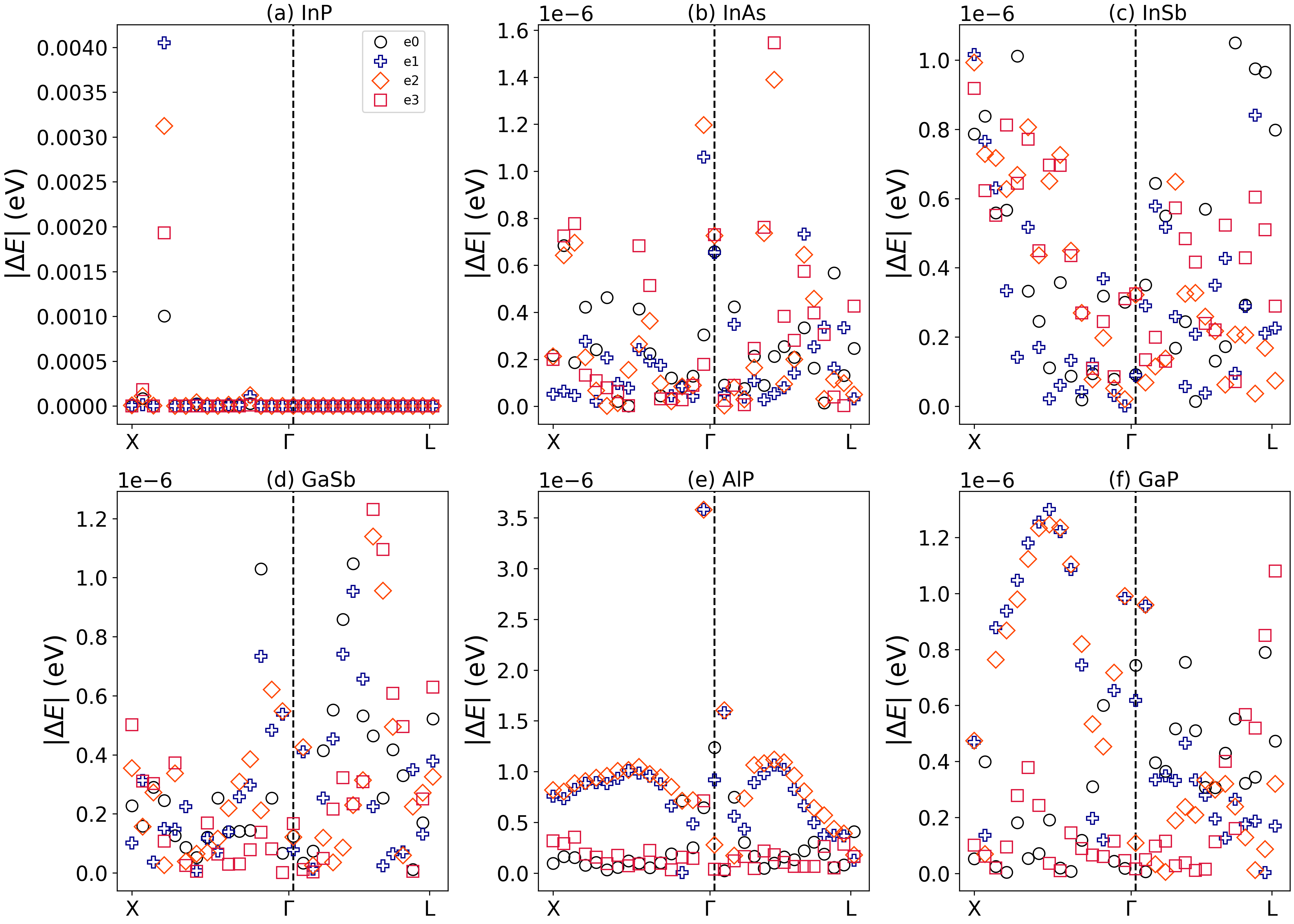}
    \caption{Energy difference for several III-V family compounds, simulated using the state-vector solver.}
    \label{fig:my_label}
\end{figure}

\begin{figure}[htb!]
    \centering
    \includegraphics[scale=0.35]{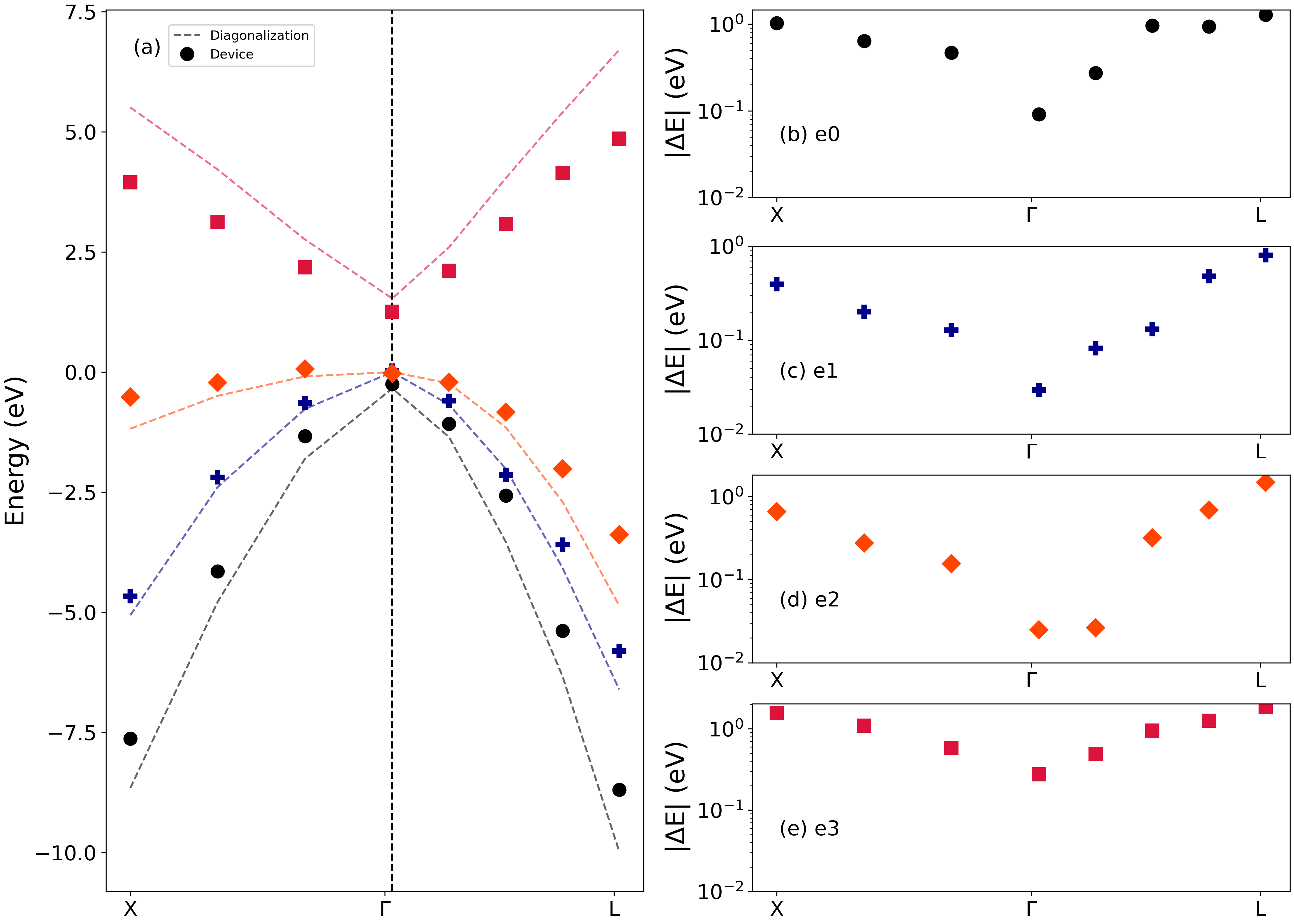}
    \caption{(a) GaAs band structure evaluated using \textit{ibm-lima} device simulator (symbols) and numerical diagonalization (dashed lines). From (b) to (e) we have the energy difference for each band.}
    \label{fig:gaasPennySim}
\end{figure}

\bibliographystyle{plain}
\bibliography{refs.bib}